\documentclass[twocolumn]{aastex62}

\usepackage{amsmath}

\shorttitle{SDSS~J1228 Thermohaline Mixing}
\shortauthors{Dwomoh \& Bauer}

\begin{document}

\title{Reinterpreting the Polluted White Dwarf SDSS~J122859.93+104032.9 in Light of Thermohaline Mixing Models: \\
More Polluting Material from a Larger Orbiting Solid Body}

\correspondingauthor{Arianna M.\ Dwomoh}
\email{arianna.dwomoh@duke.edu}

\author[0000-0002-0800-7894]{Arianna M.\ Dwomoh}
\affil{Duke University,
 2138 Campus Drive,
Durham, NC 27708, USA}
\affiliation{Center for Astrophysics $|$ Harvard \& Smithsonian,
60 Garden Street,
Cambridge, MA 02138, USA}

\author[0000-0002-4791-6724]{Evan B.~Bauer}
\affiliation{Center for Astrophysics $|$ Harvard \& Smithsonian,
60 Garden Street,
Cambridge, MA 02138, USA}

\begin{abstract}
The polluted white dwarf (WD) system SDSS~J122859.93+104032.9 (SDSS~J1228) shows variable emission features interpreted as originating from a solid core fragment held together against tidal forces by its own internal strength, orbiting within its surrounding debris disk. Estimating the size of this orbiting solid body requires modeling the accretion rate of the polluting material that is observed mixing into the WD surface. That material is supplied via sublimation from the surface of the orbiting solid body. The sublimation rate can be estimated as a simple function of the surface area of the solid body and the incident flux from the nearby hot WD.  On the other hand, estimating the accretion rate requires detailed modeling of the surface structure and mixing in the accreting WD. In this work, we present MESA WD models for SDSS~J1228 that account for thermohaline instability and mixing in addition to heavy element sedimentation to accurately constrain the sublimation and accretion rate necessary to supply the observed pollution. We derive a total accretion rate of $\dot M_{\rm acc}=1.8\times 10^{11}\,\rm g\,s^{-1}$, several orders of magnitude higher than the $\dot M_{\rm acc}=5.6\times 10^{8}\,\rm g\,s^{-1}$ estimate obtained in earlier efforts. The larger mass accretion rate implies that the minimum estimated radius of the orbiting solid body is r$_{\rm{min}}$ = 72\,km, which, although significantly larger than prior estimates, still lies within upper bounds (a few hundred km) for which the internal strength could no longer withstand tidal forces from the gravity of the WD.
\end{abstract}


\section{Introduction} \label{sec:intro}
A white dwarf (WD) is left behind when low- and intermediate-mass stars ($M \lesssim 8\, M_\odot$) reach the last stages of their evolution \citep{Fontaine2001}. WDs are very compact and have strong gravitational fields, which can tidally disrupt planetary bodies when they approach within $\sim 1 \, R_\odot$ of a WD. They then form an accreting debris disk centered on the WD, and the signatures of accretion can furnish surprising clues about the objects that were made up of that shredded material \citep{Jura2003, JuraYoung2014, Veras2016, Farihi2016}. 

Because elements heavier than helium quickly sink below the WD surface \citep{Schatzman1945}, heavy elements observed at the surface of a WD are known as pollution because they must be continuously supplied by an external source to be observable. Approximately 30\% of WDs show evidence of pollution \citep{Koester2014}, and a significant fraction of those show infrared excesses corresponding to a debris disk \citep{Farihi2016}. Recent discoveries, such as transits from orbiting debris around polluted WDs \citep{Vanderburg2015,Vanderbosch2020,Vanderbosch2021,Guidry2021}, have corroborated the emerging theoretical picture of polluted WDs supplied by disrupted planetary bodies. 

Comprehensive modeling of the chemical mixing processes at the surface of WDs allows for inferences of compositions and accretion rates supplied by disrupted planetary bodies \citep{Koester2009, Bauer2018, Bauer2019}. Such models allow us to gain insight into the chemical composition of planetary bodies, which can be utilized to study the planetary systems that once surrounded the star \citep{Zuckerman2011, JuraYoung2014, Manser2016, Manser2019, Trierweiler2022}. 

In the present work, we examine how accreted elements are mixed into the surface layers of WD models, and compare to observed spectra for the WD SDSS~J122859.93+104032.9 (hereafter SDSS~J1228). SDSS~J1228 is a WD that exhibits signatures of both photospheric pollution and infrared excess from a surrounding debris disk \citep{Gansicke2012}. The system also exhibits Ca II emission features indicating a planetesimal orbiting the WD, embedded within the debris disk inside its Roche radius \citep{Manser2019}. This implies that the planetesimal is held together by its own internal strength, preventing it from being tidally disrupted by the WD's gravity. This planetesimal may be the core of a larger differentiated object that was originally stripped of its crust and mantle (e.g., \citealt{Bonsor2020, Buchan2022, Brouwers2023}).

The accretion of heavy elements onto an atmosphere dominated by hydrogen can induce thermohaline instability that leads to extra mixing of accreted material \citep{Deal2013, Vauclair2015, Brassard2015}. It mixes on a very short timescale compared to particle diffusion, and as the surface abundances reach a steady state, mixing of accreted material reaches greater depths \citep{Wachlin2017}. Recent work suggests that in hydrogen-rich WDs with thin or no surface convection zones ($T_{\rm eff} \gtrsim 12,000\,\rm K$), the accretion rates needed in order to reproduce observed element abundances exceed those calculated without accounting for thermohaline mixing by up to three orders of magnitude \citep{Bauer2018, Bauer2019, Wachlin2022}. For SDSS~J1228 in particular, thermohaline mixing is relevant because the WD surface temperature is high enough that it should have no surface convection zone. This means that accreted heavy elements concentrate at the surface, which in turn excites the thermohaline instability.

 Here we use MESA to create models of the WD surface of SDSS~J1228 that account for thermohaline mixing. We find that previous inferences of its accretion rate are orders of magnitude too small. This implies that the solid planetary body supplying the polluting material is significantly larger than previously inferred, because its size influences the total sublimation rate responsible for supplying the accreting material to the WD. 
 
This paper is structured as follows. Sec.~\ref{sec:wd intro} presents the observed properties of SDSS~J1228 that match our models. Sec.~\ref{sec:models} describes the MESA calculations and the results obtained from the new models. In Sec.~\ref{sec:results}, we conclude by stating our findings and comparing them to previous work that did not account for thermohaline mixing.

\section{Observations of SDSS~J1228}\label{sec:wd intro}
 We build models consistent with \cite{Manser2016, Manser2019} for the WD surface structure and chemical elements observed at the surface, which are based on the observations of \cite{Gansicke2012} and \cite{Koester2014} to constrain its mass, temperature, and composition. This information is summarized in Tables~\ref{tab:observables} and~\ref{tab:observables2}.

\begin{deluxetable}{c|CC}
\tablecaption{Measured Elemental Abundances for SDSS~J1228 \label{tab:observables}}
\tablenum{1}
\tablehead{
\colhead{Element} &
\colhead{$\log(n_i/n_H)$} &
\colhead{Mass Fraction}
}
\startdata
\rm C & -7.50 \pm 0.2 & 3.79 \times 10^{-7} \\
\rm O & -4.55 \pm 0.2 & 4.51 \times 10^{-4} \\
\rm Mg & -5.10 \pm 0.2 & 1.91 \times 10^{-4} \\
\rm Al & -5.75 \pm 0.2 & 4.80 \times 10^{-5} \\
\rm Si & -5.20 \pm 0.2 & 1.77 \times 10^{-4} \\
\rm Ca & -5.70 \pm 0.2 & 7.98 \times 10^{-5} \\
\rm Fe & -5.20 \pm 0.3 & 3.53 \times 10^{-4} \\
\enddata
\tablecomments{From \cite{Gansicke2012}.}
\end{deluxetable}

\begin{deluxetable}{cCcC}
\tablecaption{Adopted Stellar Properties of SDSS~J1228 \label{tab:observables2}}
\tablenum{2}
\tablehead{
\colhead{$T_{\rm eff}\: [\rm K]$} &
\colhead{$M_{\rm WD}\: [\rm M_\odot]$} & 
\colhead{$R_{\rm WD}\: [\rm R_\odot]$} & 
\colhead{$\log(g/{\rm cm\,s^{-2}})$}
}
\startdata
$20713 \pm 281$ & $0.705 \pm 0.050$ & 0.01169 & $8.150 \pm 0.04$ \\
\enddata
\tablecomments{From \cite{Koester2014, Gansicke2012, Manser2016, Manser2019}.}
\end{deluxetable}

\section{Theoretical Models} \label{sec:models}
In this section we describe the computational models used to determine WD accretion rates, accounting for thermohaline mixing. In order to infer the interior composition and structure of the planetesimal, we build MESA models of accreting WDs representative of SDSS~J1228.

\subsection{MESA Models of SDSS~J1228}
We employ the open-source stellar evolution code MESA, version r22.05.1 \citep{Paxton2011,Paxton2013, Paxton2015, Paxton2018, Paxton2019, Jermyn2022}.
The MESA equation of state (EOS) is a blend of the OPAL \citep{Rogers2002}, SCVH
\citep{Saumon1995}, FreeEOS \citep{Irwin2004}, HELM \citep{Timmes2000},
PC \citep{Potekhin2010}, and Skye \citep{Jermyn2021} EOSes.
Radiative opacities are primarily from OPAL \citep{Iglesias1993,
Iglesias1996}.  Electron conduction opacities are from
\citet{Cassisi2007} and \cite{Blouin2020}.
Thermal neutrino loss rates are from \citet{Itoh1996}.
A repository of work directories containing MESA input files needed to reproduce all of the models presented in this work is available on Zenodo: \dataset[doi:10.5281/zenodo.7996400]{\doi{10.5281/zenodo.7996400}}.

\subsubsection{Template Model}
In order to create a MESA model representative of SDSS~J1228, we began by building a 0.705~$M_\odot$ WD model with the \texttt{make\_co\_wd} test case template in MESA, and then cooled the WD to the observed temperature of 20,713 K \citep{Gansicke2012}. Element diffusion was enabled from the beginning of the WD cooling track, to allow the WD atmosphere to stratify and the surface composition of the model to develop to a pure hydrogen composition before accretion begins. We used this cooled template WD as a starting point for all subsequent modeling.

In order to calculate the accretion rates of each element for models without thermohaline mixing, we first calculate the diffusion timescales and observed mass fractions in the photosphere. Element diffusion velocities and composition changes in MESA are calculated using an approach based on \cite{Burgers1969} with diffusion coefficients based on \cite{Stanton2016} and \cite{Caplan2022} (see \citealt{Paxton2018} for more details). Following the approach of \cite{Bauer2018}, the diffusion timescale for element species $i$ is
\begin{equation}
\tau_{{\rm diff}, i} = \frac{M_{\rm phot}}{4\pi r^2 \rho v_{{\rm diff}, i}}~.
\end{equation} 
We use the MESA WD model to calculate the density ($\rho$), surface mass ($M_{\rm phot}$), radius ($\rm{r}$), and downward sedimentation velocity ($v_{{\rm diff}, i}$). \cite{Bauer2018} employed $M_{\rm cvz}$ (mass contained in the fully-mixed surface convection zone) as opposed to $M_{\rm phot}$ for calculating the diffusion timescale, but our WD model has no surface convection zone due to its high temperature, so we evaluate the mass and diffusion at the photosphere of the model. The mass fraction of an accreted element that should be reached in diffusive equilibrium is
\begin{equation}
X_{{\rm eq}, i} = A_i \times 10^{\log(n_i/n_H)}
\end{equation}
where $A_i$ is the atomic weight for each element and $n_i$ is the observed element number density. All elements considered here are listed in Table~\ref{tab:observables}. 

Now we can find the accretion rate for each element (assuming no mixing other than element diffusion is present), which we compare to the thermohaline models later on. Solving Equation~(4) of \cite{Bauer2018} for total accretion rate
\begin{equation} 
\label{eq:MdotEq}
\dot M_i = \frac{X_{{\rm eq},i} M_{\rm phot}}{\tau_{{\rm diff},i}}~, 
\end{equation} 
we use the previously calculated surface mass, mass fraction, and diffusion timescale to find $\dot{M}_{i}$ for each element. Our results are shown in Table~\ref{tab:diffusion_results}. When comparing our calculated accretion rates to those in table 4 from \cite{Gansicke2012}, we find that our results are within a factor of 2.
Our total accretion rate for models without thermohaline mixing is $3.9 \times 10^8 \, \rm g\,s^{-1}$, which is slightly smaller than the value of $5.6 \times 10^8 \, \rm g\,s^{-1}$ derived by \cite{Gansicke2012} but well within observational uncertainties.

\begin{deluxetable}{C|CC}
\tablecaption{Diffusion timescales and accretion rates calculated assuming only element diffusion is present at the surface of our MESA models. \label{tab:diffusion_results}}
\tablenum{3}
\tablehead{
\colhead{Element} & \colhead{$\log(\tau_{\rm diff}/{\rm year})$} & \colhead{$\dot{M}_{\rm diff}\: [\rm g\,s^{-1}]$}
}
\startdata
\rm C & -2.04 & 3.12 \times 10^{4} \\
\rm O & -2.67 & 1.58 \times 10^{8} \\
\rm Mg & -2.35 & 3.21 \times 10^{7} \\
\rm Al & -2.38 & 8.51 \times 10^{6} \\
\rm Si & -2.42 & 3.42 \times 10^{7} \\
\rm Ca & -2.58 & 2.26 \times 10^{7} \\
\rm Fe & -2.72 & 1.37 \times 10^{8} \\
\enddata
\end{deluxetable}

\begin{figure*}
\plottwo{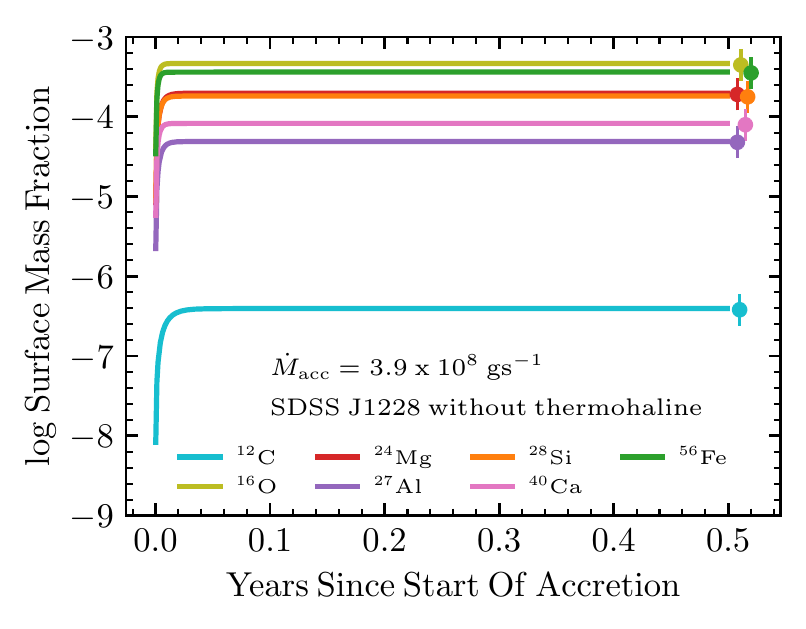}{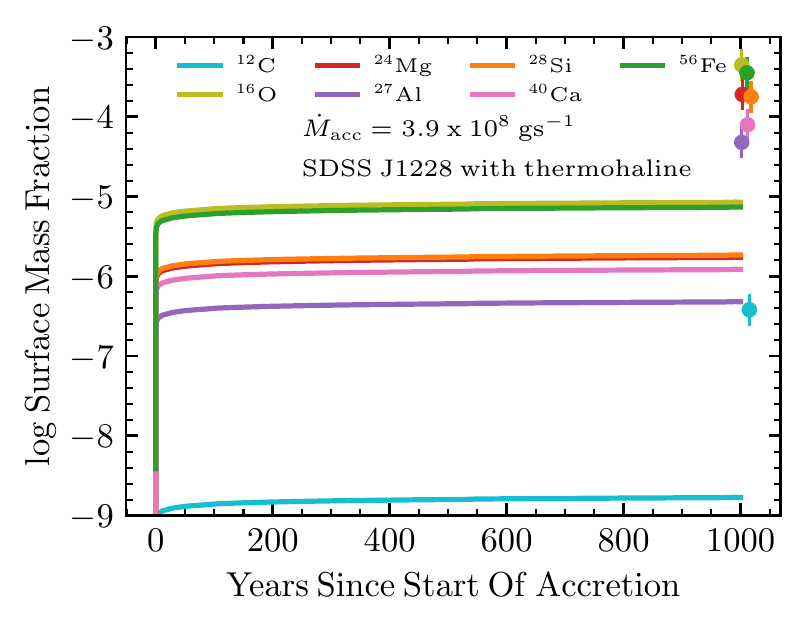}
\caption{Time evolution of surface mass fractions at a set accretion rate. {\it Left}: A MESA model of SDSS~J1228 with a calculated accretion rate of $\dot{M} = 3.9 \times 10^8\,\rm g\,s^{-1}$. No thermohaline mixing is included in this model. {\it Right}: A MESA model of the system, this time including thermohaline mixing. The steady-state surface abundances are much lower in the case where thermohaline mixing is included. The dots on both figures represent the observed surface mass fractions (listed in Table~\ref{tab:observables}) that our models must reach, with approximate error bars of 0.2~dex based on \cite{Gansicke2012}.} \label{fig:first therm no therm comparison}
\end{figure*}

\subsubsection{Steady-State Model}
\label{s.steady-state}

In order to verify that MESA models reach the expected steady state after accreting for many diffusion timescales in the absence of any mixing other than element diffusion, we run a MESA model using the accretion rates calculated using Equation~\eqref{eq:MdotEq}. This model allows us to verify that in steady state the surface mass fractions match the observed surface composition for each element. In the left-hand panel of Figure~\ref{fig:first therm no therm comparison}, the steady state matches observed mass fractions when only diffusion is accounted for. We define a ``match'' to be when the surface mass fractions produced by MESA are within 10\% of the \cite{Gansicke2012} observations. 

As a further test, we then run a model with the same accretion rate with thermohaline mixing turned on, which is shown in the right panel of Figure~\ref{fig:first therm no therm comparison}. When thermohaline mixing is included in the model, the mass fractions at the surface of the model decrease by roughly two orders of magnitude for the same accretion rate. This shows that thermohaline mixing dramatically alters the observed surface abundance by mixing accreted material further into the WD. It also provides an initial estimate for how much greater the modeled accretion rate will need to be to match the observed surface abundances.

\subsubsection{Thermohaline Mixing}

Thermohaline instability can drive mixing in fluids that have stable temperature stratification but unstable composition gradients.
For example, thermohaline mixing occurs in the oceans in regions where warm saltwater is above a layer of cool, less salty water, forming downward sinking ``fingers" when the saltwater begins to cool off \citep{BainesGill1969}. Stars accreting planetary debris can experience a similar instability due to mixing heavy elements into surfaces composed of primarily hydrogen or helium \citep{Kippenhahn1980, Garaud2018, Garaud2021, Sevilla2022, Behmard2023}. 
Molecular weight gradients can cause fluid instability in stars, and for some polluted WDs, drive the thermohaline instability \citep{Deal2013,Bauer2019,Wachlin2022}.

The strength of mixing due to thermohaline instability can be understood in terms of a diffusion coefficient $D_{\rm th}$, which scales approximately as
\begin{equation}
    D_{\rm th} \propto \kappa_T \frac{\nabla_\mu}{\nabla_T - \nabla_{\rm ad}}~,
    \label{eq:Dth}
\end{equation}
and instability is present when $\nabla_T - \nabla_{\rm ad} < \nabla_\mu < 0$
\citep{Kippenhahn1980}.
In the equation above, $\nabla_\mu, \nabla_T, \nabla_{\rm ad}$ and $\kappa_T$ represent the mean molecular weight gradient, temperature gradient in the fluid, adiabatic temperature gradient, and thermal diffusivity, respectively.

We build MESA models that include thermohaline mixing according to the prescription of \cite{Brown2013}.%
\footnote{This prescription employs the notion of ``parasitic saturation'' to enable estimating the total amount of mixing in 1D models by calculating when the mixing will saturate due to secondary shear instabilities in the fingers. It is therefore somehwat more sophisticated than the simplified model presented in Eqn~\eqref{eq:Dth}, but it has been shown to produce net mixing that scales similarly in the polluted WD context \citep{Bauer2019}. The prescription of \cite{Brown2013} has been extensively validated against 3D simulations in the hydrodynamical regime, though 3D magnetohydrodynamical simulations have shown that thermohaline mixing could be further enhanced in the presence of magnetic fields \citep{Harrington2019,Fraser2023}.}
When running models that include both thermohaline mixing and element diffusion, we verify that diffusion no longer has a noticeable effect on the predicted surface abundances. Because thermohaline mixing dominates when both processes are present, we ignore diffusion for our subsequent models, and focus on models that include only thermohaline mixing to find the accretion rate necessary to match observations.

Increasing the accretion rate leads to greater thermohaline instability and more mixing, so the amount of accretion needed to match the particular amount of pollution is not a simple linear function. We tune our models to match observations by iteratively adjusting the accretion rate and checking how the steady-state surface abundances compare to observed values.

\begin{figure}
\plotone{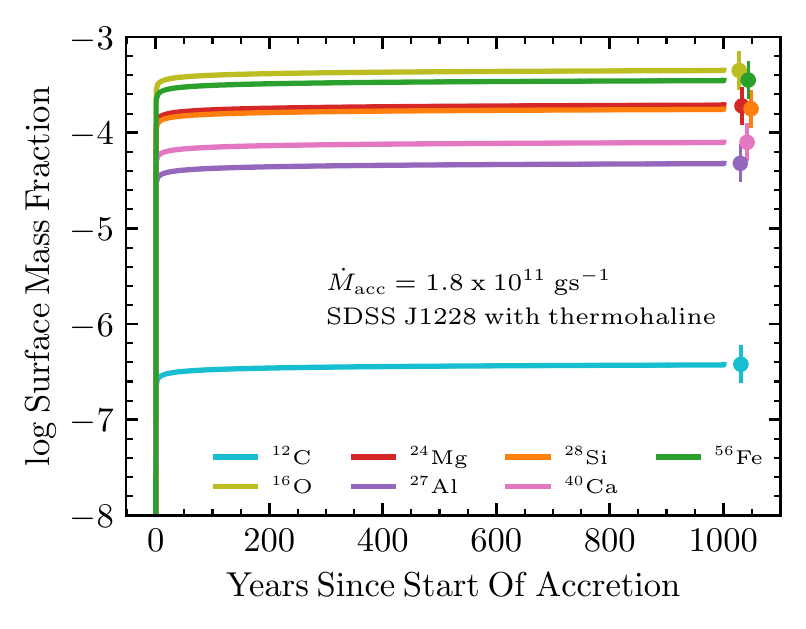}
\caption{Time evolution of the thermohaline model with $\dot{M}$ tuned to match observed mass fractions over a period of 1000 years. The dots represent the observed surface mass fractions (listed in Table~\ref{tab:observables}) that our models much reach, with approximate error bars of 0.2~dex.  \label{fig:therm model closer to observed}}
\end{figure}

\begin{figure}
\plotone{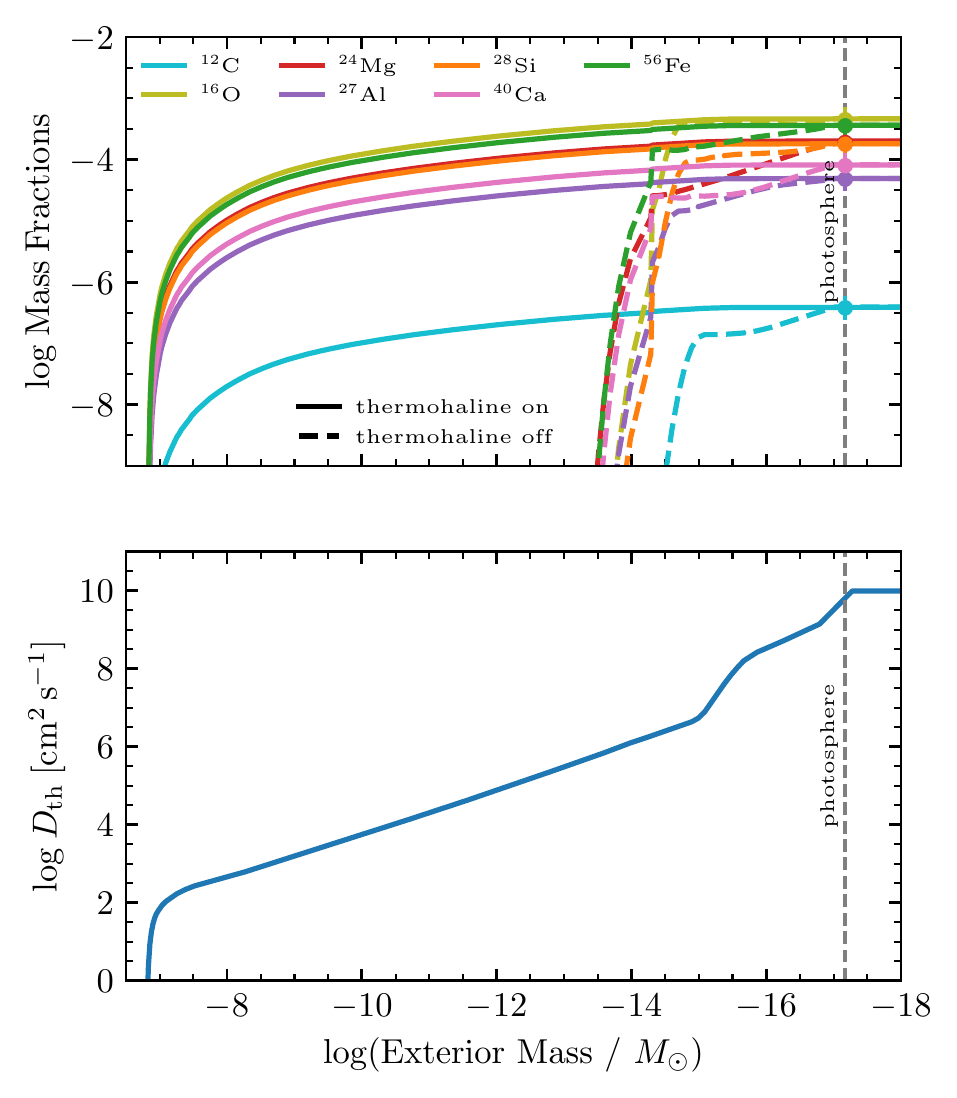} 
\caption{{\it Upper:} Mass fraction profiles of the accreted polluting elements mixing into the surface layers of the WD model. {\it Lower:} Thermohaline mixing coefficient profile.
The surface of the WD lies at the right on the x-axis, while deeper layers lie toward the left.
The dashed lines in the upper panel represent the composition profile for the MESA model that matches SDSS~J1228 without thermohaline mixing from the left panel of Figure~\ref{fig:first therm no therm comparison}, and the dots represent the observed composition at the photosphere.
\label{fig:diff timescale for therm}}
\end{figure}

Our results after tuning $\dot M$ to match SDSS~J1228 are shown in Figure~\ref{fig:therm model closer to observed}. Figure~\ref{fig:diff timescale for therm} shows the interior abundance and mixing profile for the same MESA model, along with the composition profile from the model without thermohaline mixing from Section~\ref{s.steady-state} for comparison. The presence of thermohaline mixing, as compared to diffusion, results in accreted material being mixed much deeper into the WD on a faster timescale. Thus, the accretion rate must be higher in order to continue matching the observed surface abundances.

\newpage
\section{Results and Conclusion} \label{sec:results}

\subsection{Accretion Rate and Composition}
When including thermohaline mixing, our MESA modeling of SDSS~J1228 finds that the best match for the accretion rate is $1.8 \times 10^{11}\,\rm g\,s^{-1}$ $(2.8 \times 10^{-15}\, \rm M_\odot\,yr^{-1})$, more than two orders of magnitude higher than previously inferred for this object. Table~\ref{tab:data from analysis} shows the changes in total accretion rate and other inferred properties when comparing our MESA models with and without thermohaline mixing.

We also find that the relative accreted mass fractions of material in the WD photosphere are different when accounting for thermohaline mixing. Notably, the accreted fractions of $^{16}$O and $^{56}$Fe are somewhat lower than previously inferred, and the accreted fractions of $^{24}$Mg and $^{28}$Si are significantly higher (see Table~\ref{tab:data from analysis}).

\begin{deluxetable*}{C|cc} 
\tablecaption{Comparison of accreted material and parent body properties in SDSS~J1228 inferred from models with and without the presence of thermohaline mixing \label{tab:data from analysis}}
\tablenum{4}
\tablehead{
\colhead{} & \colhead{With Thermohaline Mixing} & \colhead{Without Thermohaline Mixing}
}
\startdata
$\dot{M}_{\rm acc}\: \rm{[g\,s^{-1}]}$ & $1.8 \times 10^{11}$ & $3.9 \times 10^8$ \\  
$r_{\rm min}\: \rm{[km]}$ & 72 & 4 \\ 
$\rm{lifetime\: [yr]}$ & 1500 & 85 \\
$\rm{total\: mass\: [g]}$ & $1.3 \times 10^{22}$ & $2.1 \times 10^{18}$ \\ 
$^{16}{\rm O\: Mass\: Fraction}$ & $0.347$ & $0.402$ \\ 
$^{24}\rm{Mg\: Mass\: Fraction}$ & $0.151$ & $0.082$ \\ 
$^{27}\rm{Al\: Mass\: Fraction}$ & $0.037$ & $0.022$ \\
$^{28}\rm{Si\: Mass\: Fraction}$ & $0.136$ & $0.087$ \\ 
$^{40}\rm{Ca\: Mass\: Fraction}$ & $0.061$ & $0.058$ \\ 
$^{56}\rm{Fe\: Mass\: Fraction}$ & $0.272$ & $0.349$ \\ 
\enddata
\end{deluxetable*}

\subsection{Size of the Parent Body}

\cite{Manser2019} have argued that the pollution and gas observed in SDSS~J1228 are supplied by a solid iron-rich body with internal strength orbiting within the Roche radius of the WD. Based on the density of iron and internal strength of iron meteorite samples, they estimate an upper limit of a few hundred km for the size $r$ of the solid object before tidal forces from the WD gravity would overcome the solid body's internal strength and coherence. They estimate a lower limit for the size $r$ by arguing that the observed polluting material is supplied by evaporation of the solid body due to irradiation from the nearby WD. The surface area of the solid body sets the amount of irradiation energy from the WD incident on the body, so the evaporation rate scales as $\dot M \propto r^2$, i.e.\ minimum inferred size scales as $r_{\rm min} \propto \sqrt{\dot M}$. Based on the calculated accretion rate of $\dot M = 5.6 \times 10^8\, \rm g\,s^{-1}$ from \cite{Gansicke2012}, \cite{Manser2019} infer an approximate minimum size of $r_{\rm min}\approx 4\,\rm km$ for the solid body orbiting SDSS~J1228. With this size and a density comparable to iron ($\approx 8\, \rm g\,cm^{-3}$), the total lifetime of the solid body would be just $\approx 85\,\rm yr$ before its entire mass evaporates away.

We reassess this lower limit with our calculated accretion rate of $1.8 \times 10^{11} \,\rm g\,s^{-1}$ accounting for thermohaline mixing. Due to this higher accretion rate, the minimum size needed to supply this accretion from evaporation is $r_{\rm min} \approx 72\, \rm km$, and the corresponding lifetime is $\approx 1500\,\rm yr$. This lower limit still lies within the upper limit of a few hundred km from considering internal strength vs tidal forces, but significantly narrows the overall range of possible sizes for the solid body. It also points to a much more plausible, longer lifespan for the currently observed phase of the evaporating solid body. This longer timescale ($1500\,\rm yr$ instead of just $85\,\rm yr$) before the orbiting mass is sublimated makes it much more likely that human observations could catch a polluted WD in a state like that observed for SDSS~J1228.

\subsection{Conclusion}
By incorporating an updated physical model for the surface of the accreting WD in SDSS~J1228, our work has yielded a better understanding of the planetesimal orbiting SDSS~J1228. Our models lead to an inferred accretion rate of $1.8 \times 10^{11}\,\rm g\,s^{-1}$, more than two orders of magnitude higher than previously inferred. This accretion rate in turn implies a much large minimum size and mass of the inferred solid core fragment orbiting in the debris disk of SDSS~J1228. Future studies of warm polluted DA WDs, such as the one observed in SDSS~J1228, should account for thermohaline mixing when making model-based inferences about accretion properties.

\acknowledgments
{\it Acknowledgments:} We thank the anonymous referee for constructive feedback on an earlier draft of this work. We would like to thank Jonathan McDowell and Matthew Ashby for serving as amazing mentors through the SAO REU program, and for their continued support and assistance during the paper-writing process. 

The SAO REU program is funded in part by the National Science Foundation REU and Department of Defense ASSURE programs under NSF Grant no. AST-2050813, and by the Smithsonian Institution.

\vspace{5mm}

\software{Astropy \citep{astropy:2013, astropy:2018, astropy:2022}, Matplotlib \citep{Hunter2007}, Modules for Experiments in Stellar Astrophysics (MESA, \citealt{Paxton2011, Paxton2013, Paxton2015, Paxton2018, Paxton2019, Jermyn2022}).}

\bibliographystyle{aasjournal}
\bibliography{paper,mesa}

\begin{thebibliography}{}
\expandafter\ifx\csname natexlab\endcsname\relax\def\natexlab#1{#1}\fi
\providecommand{\url}[1]{\href{#1}{#1}}
\providecommand{\dodoi}[1]{doi:~\href{http://doi.org/#1}{\nolinkurl{#1}}}
\providecommand{\doeprint}[1]{\href{http://ascl.net/#1}{\nolinkurl{http://ascl.net/#1}}}
\providecommand{\doarXiv}[1]{\href{https://arxiv.org/abs/#1}{\nolinkurl{https://arxiv.org/abs/#1}}}

\bibitem[{{Astropy Collaboration} {et~al.}(2013){Astropy Collaboration},
  {Robitaille}, {Tollerud}, {Greenfield}, {Droettboom}, {Bray}, {Aldcroft},
  {Davis}, {Ginsburg}, {Price-Whelan}, {Kerzendorf}, {Conley}, {Crighton},
  {Barbary}, {Muna}, {Ferguson}, {Grollier}, {Parikh}, {Nair}, {Unther},
  {Deil}, {Woillez}, {Conseil}, {Kramer}, {Turner}, {Singer}, {Fox}, {Weaver},
  {Zabalza}, {Edwards}, {Azalee Bostroem}, {Burke}, {Casey}, {Crawford},
  {Dencheva}, {Ely}, {Jenness}, {Labrie}, {Lim}, {Pierfederici}, {Pontzen},
  {Ptak}, {Refsdal}, {Servillat}, \& {Streicher}}]{astropy:2013}
{Astropy Collaboration}, {Robitaille}, T.~P., {Tollerud}, E.~J., {et~al.} 2013,
  \aap, 558, A33, \dodoi{10.1051/0004-6361/201322068}

\bibitem[{{Astropy Collaboration} {et~al.}(2018){Astropy Collaboration},
  {Price-Whelan}, {Sip{\H{o}}cz}, {G{\"u}nther}, {Lim}, {Crawford}, {Conseil},
  {Shupe}, {Craig}, {Dencheva}, {Ginsburg}, {Vand erPlas}, {Bradley},
  {P{\'e}rez-Su{\'a}rez}, {de Val-Borro}, {Aldcroft}, {Cruz}, {Robitaille},
  {Tollerud}, {Ardelean}, {Babej}, {Bach}, {Bachetti}, {Bakanov}, {Bamford},
  {Barentsen}, {Barmby}, {Baumbach}, {Berry}, {Biscani}, {Boquien}, {Bostroem},
  {Bouma}, {Brammer}, {Bray}, {Breytenbach}, {Buddelmeijer}, {Burke},
  {Calderone}, {Cano Rodr{\'\i}guez}, {Cara}, {Cardoso}, {Cheedella}, {Copin},
  {Corrales}, {Crichton}, {D'Avella}, {Deil}, {Depagne}, {Dietrich}, {Donath},
  {Droettboom}, {Earl}, {Erben}, {Fabbro}, {Ferreira}, {Finethy}, {Fox},
  {Garrison}, {Gibbons}, {Goldstein}, {Gommers}, {Greco}, {Greenfield},
  {Groener}, {Grollier}, {Hagen}, {Hirst}, {Homeier}, {Horton}, {Hosseinzadeh},
  {Hu}, {Hunkeler}, {Ivezi{\'c}}, {Jain}, {Jenness}, {Kanarek}, {Kendrew},
  {Kern}, {Kerzendorf}, {Khvalko}, {King}, {Kirkby}, {Kulkarni}, {Kumar},
  {Lee}, {Lenz}, {Littlefair}, {Ma}, {Macleod}, {Mastropietro}, {McCully},
  {Montagnac}, {Morris}, {Mueller}, {Mumford}, {Muna}, {Murphy}, {Nelson},
  {Nguyen}, {Ninan}, {N{\"o}the}, {Ogaz}, {Oh}, {Parejko}, {Parley}, {Pascual},
  {Patil}, {Patil}, {Plunkett}, {Prochaska}, {Rastogi}, {Reddy Janga},
  {Sabater}, {Sakurikar}, {Seifert}, {Sherbert}, {Sherwood-Taylor}, {Shih},
  {Sick}, {Silbiger}, {Singanamalla}, {Singer}, {Sladen}, {Sooley},
  {Sornarajah}, {Streicher}, {Teuben}, {Thomas}, {Tremblay}, {Turner},
  {Terr{\'o}n}, {van Kerkwijk}, {de la Vega}, {Watkins}, {Weaver}, {Whitmore},
  {Woillez}, {Zabalza}, \& {Astropy Contributors}}]{astropy:2018}
{Astropy Collaboration}, {Price-Whelan}, A.~M., {Sip{\H{o}}cz}, B.~M., {et~al.}
  2018, \aj, 156, 123, \dodoi{10.3847/1538-3881/aabc4f}

\bibitem[{{Astropy Collaboration} {et~al.}(2022){Astropy Collaboration},
  {Price-Whelan}, {Lim}, {Earl}, {Starkman}, {Bradley}, {Shupe}, {Patil},
  {Corrales}, {Brasseur}, {N{\"o}the}, {Donath}, {Tollerud}, {Morris},
  {Ginsburg}, {Vaher}, {Weaver}, {Tocknell}, {Jamieson}, {van Kerkwijk},
  {Robitaille}, {Merry}, {Bachetti}, {G{\"u}nther}, {Aldcroft},
  {Alvarado-Montes}, {Archibald}, {B{\'o}di}, {Bapat}, {Barentsen},
  {Baz{\'a}n}, {Biswas}, {Boquien}, {Burke}, {Cara}, {Cara}, {Conroy},
  {Conseil}, {Craig}, {Cross}, {Cruz}, {D'Eugenio}, {Dencheva}, {Devillepoix},
  {Dietrich}, {Eigenbrot}, {Erben}, {Ferreira}, {Foreman-Mackey}, {Fox},
  {Freij}, {Garg}, {Geda}, {Glattly}, {Gondhalekar}, {Gordon}, {Grant},
  {Greenfield}, {Groener}, {Guest}, {Gurovich}, {Handberg}, {Hart},
  {Hatfield-Dodds}, {Homeier}, {Hosseinzadeh}, {Jenness}, {Jones}, {Joseph},
  {Kalmbach}, {Karamehmetoglu}, {Ka{\l}uszy{\'n}ski}, {Kelley}, {Kern},
  {Kerzendorf}, {Koch}, {Kulumani}, {Lee}, {Ly}, {Ma}, {MacBride}, {Maljaars},
  {Muna}, {Murphy}, {Norman}, {O'Steen}, {Oman}, {Pacifici}, {Pascual},
  {Pascual-Granado}, {Patil}, {Perren}, {Pickering}, {Rastogi}, {Roulston},
  {Ryan}, {Rykoff}, {Sabater}, {Sakurikar}, {Salgado}, {Sanghi}, {Saunders},
  {Savchenko}, {Schwardt}, {Seifert-Eckert}, {Shih}, {Jain}, {Shukla}, {Sick},
  {Simpson}, {Singanamalla}, {Singer}, {Singhal}, {Sinha}, {Sip{\H{o}}cz},
  {Spitler}, {Stansby}, {Streicher}, {{\v{S}}umak}, {Swinbank}, {Taranu},
  {Tewary}, {Tremblay}, {de Val-Borro}, {Van Kooten}, {Vasovi{\'c}}, {Verma},
  {de Miranda Cardoso}, {Williams}, {Wilson}, {Winkel}, {Wood-Vasey}, {Xue},
  {Yoachim}, {Zhang}, {Zonca}, \& {Astropy Project
  Contributors}}]{astropy:2022}
{Astropy Collaboration}, {Price-Whelan}, A.~M., {Lim}, P.~L., {et~al.} 2022,
  \apj, 935, 167, \dodoi{10.3847/1538-4357/ac7c74}

\bibitem[{{Baines} \& {Gill}(1969)}]{BainesGill1969}
{Baines}, P.~G., \& {Gill}, A.~E. 1969, Journal of Fluid Mechanics, 37, 289,
  \dodoi{10.1017/S0022112069000553}

\bibitem[{{Bauer} \& {Bildsten}(2018)}]{Bauer2018}
{Bauer}, E.~B., \& {Bildsten}, L. 2018, \apjl, 859, L19,
  \dodoi{10.3847/2041-8213/aac492}

\bibitem[{{Bauer} \& {Bildsten}(2019)}]{Bauer2019}
---. 2019, \apj, 872, 96, \dodoi{10.3847/1538-4357/ab0028}

\bibitem[{{Behmard} {et~al.}(2023){Behmard}, {Sevilla}, \&
  {Fuller}}]{Behmard2023}
{Behmard}, A., {Sevilla}, J., \& {Fuller}, J. 2023, \mnras, 518, 5465,
  \dodoi{10.1093/mnras/stac3435}

\bibitem[{{Blouin} {et~al.}(2020){Blouin}, {Shaffer}, {Saumon}, \&
  {Starrett}}]{Blouin2020}
{Blouin}, S., {Shaffer}, N.~R., {Saumon}, D., \& {Starrett}, C.~E. 2020, \apj,
  899, 46, \dodoi{10.3847/1538-4357/ab9e75}

\bibitem[{{Bonsor} {et~al.}(2020){Bonsor}, {Carter}, {Hollands},
  {G{\"a}nsicke}, {Leinhardt}, \& {Harrison}}]{Bonsor2020}
{Bonsor}, A., {Carter}, P.~J., {Hollands}, M., {et~al.} 2020, \mnras, 492,
  2683, \dodoi{10.1093/mnras/stz3603}

\bibitem[{{Brassard} \& {Fontaine}(2015)}]{Brassard2015}
{Brassard}, P., \& {Fontaine}, G. 2015, in Astronomical Society of the Pacific
  Conference Series, Vol. 493, 19th European Workshop on White Dwarfs, ed.
  P.~{Dufour}, P.~{Bergeron}, \& G.~{Fontaine}, 121

\bibitem[{{Brouwers} {et~al.}(2023){Brouwers}, {Bonsor}, \&
  {Malamud}}]{Brouwers2023}
{Brouwers}, M.~G., {Bonsor}, A., \& {Malamud}, U. 2023, \mnras, 519, 2646,
  \dodoi{10.1093/mnras/stac3316}

\bibitem[{{Brown} {et~al.}(2013){Brown}, {Garaud}, \& {Stellmach}}]{Brown2013}
{Brown}, J.~M., {Garaud}, P., \& {Stellmach}, S. 2013, \apj, 768, 34,
  \dodoi{10.1088/0004-637X/768/1/34}

\bibitem[{{Buchan} {et~al.}(2022){Buchan}, {Bonsor}, {Shorttle}, {Wade},
  {Harrison}, {Noack}, \& {Koester}}]{Buchan2022}
{Buchan}, A.~M., {Bonsor}, A., {Shorttle}, O., {et~al.} 2022, \mnras, 510,
  3512, \dodoi{10.1093/mnras/stab3624}

\bibitem[{{Burgers}(1969)}]{Burgers1969}
{Burgers}, J.~M. 1969, {Flow Equations for Composite Gases}

\bibitem[{{Caplan} {et~al.}(2022){Caplan}, {Bauer}, \& {Freeman}}]{Caplan2022}
{Caplan}, M.~E., {Bauer}, E.~B., \& {Freeman}, I.~F. 2022, \mnras, 513, L52,
  \dodoi{10.1093/mnrasl/slac032}

\bibitem[{{Cassisi} {et~al.}(2007){Cassisi}, {Potekhin}, {Pietrinferni},
  {Catelan}, \& {Salaris}}]{Cassisi2007}
{Cassisi}, S., {Potekhin}, A.~Y., {Pietrinferni}, A., {Catelan}, M., \&
  {Salaris}, M. 2007, \apj, 661, 1094, \dodoi{10.1086/516819}

\bibitem[{{Deal} {et~al.}(2013){Deal}, {Deheuvels}, {Vauclair}, {Vauclair}, \&
  {Wachlin}}]{Deal2013}
{Deal}, M., {Deheuvels}, S., {Vauclair}, G., {Vauclair}, S., \& {Wachlin},
  F.~C. 2013, \aap, 557, L12, \dodoi{10.1051/0004-6361/201322206}

\bibitem[{{Farihi}(2016)}]{Farihi2016}
{Farihi}, J. 2016, \nar, 71, 9, \dodoi{10.1016/j.newar.2016.03.001}

\bibitem[{{Fontaine} {et~al.}(2001){Fontaine}, {Brassard}, \&
  {Bergeron}}]{Fontaine2001}
{Fontaine}, G., {Brassard}, P., \& {Bergeron}, P. 2001, \pasp, 113, 409,
  \dodoi{10.1086/319535}

\bibitem[{{Fraser} \& {Garaud}(2023)}]{Fraser2023}
{Fraser}, A.~E., \& {Garaud}, P. 2023, arXiv e-prints, arXiv:2302.11610,
  \dodoi{10.48550/arXiv.2302.11610}

\bibitem[{{G{\"a}nsicke} {et~al.}(2012){G{\"a}nsicke}, {Koester}, {Farihi},
  {Girven}, {Parsons}, \& {Breedt}}]{Gansicke2012}
{G{\"a}nsicke}, B.~T., {Koester}, D., {Farihi}, J., {et~al.} 2012, \mnras, 424,
  333, \dodoi{10.1111/j.1365-2966.2012.21201.x}

\bibitem[{{Garaud}(2018)}]{Garaud2018}
{Garaud}, P. 2018, Annual Review of Fluid Mechanics, 50, 275,
  \dodoi{10.1146/annurev-fluid-122316-045234}

\bibitem[{{Garaud}(2021)}]{Garaud2021}
---. 2021, arXiv e-prints, arXiv:2103.08072.
\newblock \doarXiv{2103.08072}

\bibitem[{{Guidry} {et~al.}(2021){Guidry}, {Vanderbosch}, {Hermes}, {Barlow},
  {Lopez}, {Boudreaux}, {Corcoran}, {Bell}, {Montgomery}, {Heintz},
  {Castanheira}, {Reding}, {Dunlap}, {Winget}, {Winget}, \&
  {Kuehne}}]{Guidry2021}
{Guidry}, J.~A., {Vanderbosch}, Z.~P., {Hermes}, J.~J., {et~al.} 2021, \apj,
  912, 125, \dodoi{10.3847/1538-4357/abee68}

\bibitem[{{Harrington} \& {Garaud}(2019)}]{Harrington2019}
{Harrington}, P.~Z., \& {Garaud}, P. 2019, \apjl, 870, L5,
  \dodoi{10.3847/2041-8213/aaf812}

\bibitem[{Hunter(2007)}]{Hunter2007}
Hunter, J.~D. 2007, Computing in Science \& Engineering, 9, 90,
  \dodoi{10.1109/MCSE.2007.55}

\bibitem[{{Iglesias} \& {Rogers}(1993)}]{Iglesias1993}
{Iglesias}, C.~A., \& {Rogers}, F.~J. 1993, \apj, 412, 752,
  \dodoi{10.1086/172958}

\bibitem[{{Iglesias} \& {Rogers}(1996)}]{Iglesias1996}
---. 1996, \apj, 464, 943, \dodoi{10.1086/177381}

\bibitem[{{Irwin}(2004)}]{Irwin2004}
{Irwin}, A.~W. 2004, The FreeEOS Code for Calculating the Equation of State for
  Stellar Interiors.
\newblock \url{http://freeeos.sourceforge.net/}

\bibitem[{{Itoh} {et~al.}(1996){Itoh}, {Hayashi}, {Nishikawa}, \&
  {Kohyama}}]{Itoh1996}
{Itoh}, N., {Hayashi}, H., {Nishikawa}, A., \& {Kohyama}, Y. 1996, \apjs, 102,
  411, \dodoi{10.1086/192264}

\bibitem[{{Jermyn} {et~al.}(2021){Jermyn}, {Schwab}, {Bauer}, {Timmes}, \&
  {Potekhin}}]{Jermyn2021}
{Jermyn}, A.~S., {Schwab}, J., {Bauer}, E., {Timmes}, F.~X., \& {Potekhin},
  A.~Y. 2021, \apj, 913, 72, \dodoi{10.3847/1538-4357/abf48e}

\bibitem[{{Jermyn} {et~al.}(2023){Jermyn}, {Bauer}, {Schwab}, {Farmer}, {Ball},
  {Bellinger}, {Dotter}, {Joyce}, {Marchant}, {Mombarg}, {Wolf}, {Wong},
  {Cinquegrana}, {Farrell}, {Smolec}, {Thoul}, {Cantiello}, {Herwig}, {Toloza},
  {Bildsten}, {Townsend}, \& {Timmes}}]{Jermyn2022}
{Jermyn}, A.~S., {Bauer}, E.~B., {Schwab}, J., {et~al.} 2023, \apjs, 265, 15,
  \dodoi{10.3847/1538-4365/acae8d}

\bibitem[{{Jura}(2003)}]{Jura2003}
{Jura}, M. 2003, \apjl, 584, L91, \dodoi{10.1086/374036}

\bibitem[{{Jura} \& {Young}(2014)}]{JuraYoung2014}
{Jura}, M., \& {Young}, E.~D. 2014, Annual Review of Earth and Planetary
  Sciences, 42, 45, \dodoi{10.1146/annurev-earth-060313-054740}

\bibitem[{{Kippenhahn} {et~al.}(1980){Kippenhahn}, {Ruschenplatt}, \&
  {Thomas}}]{Kippenhahn1980}
{Kippenhahn}, R., {Ruschenplatt}, G., \& {Thomas}, H.~C. 1980, \aap, 91, 175

\bibitem[{{Koester}(2009)}]{Koester2009}
{Koester}, D. 2009, \aap, 498, 517, \dodoi{10.1051/0004-6361/200811468}

\bibitem[{{Koester} {et~al.}(2014){Koester}, {G{\"a}nsicke}, \&
  {Farihi}}]{Koester2014}
{Koester}, D., {G{\"a}nsicke}, B.~T., \& {Farihi}, J. 2014, \aap, 566, A34,
  \dodoi{10.1051/0004-6361/201423691}

\bibitem[{{Manser} {et~al.}(2016){Manser}, {G{\"a}nsicke}, {Marsh}, {Veras},
  {Koester}, {Breedt}, {Pala}, {Parsons}, \& {Southworth}}]{Manser2016}
{Manser}, C.~J., {G{\"a}nsicke}, B.~T., {Marsh}, T.~R., {et~al.} 2016, \mnras,
  455, 4467, \dodoi{10.1093/mnras/stv2603}

\bibitem[{{Manser} {et~al.}(2019){Manser}, {G{\"a}nsicke}, {Eggl}, {Hollands},
  {Izquierdo}, {Koester}, {Landstreet}, {Lyra}, {Marsh}, {Meru}, {Mustill},
  {Rodr{\'\i}guez-Gil}, {Toloza}, {Veras}, {Wilson}, {Burleigh}, {Davies},
  {Farihi}, {Gentile Fusillo}, {de Martino}, {Parsons}, {Quirrenbach}, {Raddi},
  {Reffert}, {Del Santo}, {Schreiber}, {Silvotti}, {Toonen}, {Villaver},
  {Wyatt}, {Xu}, \& {Portegies Zwart}}]{Manser2019}
{Manser}, C.~J., {G{\"a}nsicke}, B.~T., {Eggl}, S., {et~al.} 2019, Science,
  364, 66, \dodoi{10.1126/science.aat5330}

\bibitem[{{Paxton} {et~al.}(2011){Paxton}, {Bildsten}, {Dotter}, {Herwig},
  {Lesaffre}, \& {Timmes}}]{Paxton2011}
{Paxton}, B., {Bildsten}, L., {Dotter}, A., {et~al.} 2011, \apjs, 192, 3,
  \dodoi{10.1088/0067-0049/192/1/3}

\bibitem[{{Paxton} {et~al.}(2013){Paxton}, {Cantiello}, {Arras}, {Bildsten},
  {Brown}, {Dotter}, {Mankovich}, {Montgomery}, {Stello}, {Timmes}, \&
  {Townsend}}]{Paxton2013}
{Paxton}, B., {Cantiello}, M., {Arras}, P., {et~al.} 2013, \apjs, 208, 4,
  \dodoi{10.1088/0067-0049/208/1/4}

\bibitem[{{Paxton} {et~al.}(2015){Paxton}, {Marchant}, {Schwab}, {Bauer},
  {Bildsten}, {Cantiello}, {Dessart}, {Farmer}, {Hu}, {Langer}, {Townsend},
  {Townsley}, \& {Timmes}}]{Paxton2015}
{Paxton}, B., {Marchant}, P., {Schwab}, J., {et~al.} 2015, \apjs, 220, 15,
  \dodoi{10.1088/0067-0049/220/1/15}

\bibitem[{{Paxton} {et~al.}(2018){Paxton}, {Schwab}, {Bauer}, {Bildsten},
  {Blinnikov}, {Duffell}, {Farmer}, {Goldberg}, {Marchant}, {Sorokina},
  {Thoul}, {Townsend}, \& {Timmes}}]{Paxton2018}
{Paxton}, B., {Schwab}, J., {Bauer}, E.~B., {et~al.} 2018, \apjs, 234, 34,
  \dodoi{10.3847/1538-4365/aaa5a8}

\bibitem[{{Paxton} {et~al.}(2019){Paxton}, {Smolec}, {Schwab}, {Gautschy},
  {Bildsten}, {Cantiello}, {Dotter}, {Farmer}, {Goldberg}, {Jermyn}, {Kanbur},
  {Marchant}, {Thoul}, {Townsend}, {Wolf}, {Zhang}, \& {Timmes}}]{Paxton2019}
{Paxton}, B., {Smolec}, R., {Schwab}, J., {et~al.} 2019, \apjs, 243, 10,
  \dodoi{10.3847/1538-4365/ab2241}

\bibitem[{{Potekhin} \& {Chabrier}(2010)}]{Potekhin2010}
{Potekhin}, A.~Y., \& {Chabrier}, G. 2010, Contributions to Plasma Physics, 50,
  82, \dodoi{10.1002/ctpp.201010017}

\bibitem[{{Rogers} \& {Nayfonov}(2002)}]{Rogers2002}
{Rogers}, F.~J., \& {Nayfonov}, A. 2002, \apj, 576, 1064,
  \dodoi{10.1086/341894}

\bibitem[{{Saumon} {et~al.}(1995){Saumon}, {Chabrier}, \& {van
  Horn}}]{Saumon1995}
{Saumon}, D., {Chabrier}, G., \& {van Horn}, H.~M. 1995, \apjs, 99, 713,
  \dodoi{10.1086/192204}

\bibitem[{{Schatzman}(1945)}]{Schatzman1945}
{Schatzman}, E. 1945, Annales d'Astrophysique, 8, 143

\bibitem[{{Sevilla} {et~al.}(2022){Sevilla}, {Behmard}, \&
  {Fuller}}]{Sevilla2022}
{Sevilla}, J., {Behmard}, A., \& {Fuller}, J. 2022, \mnras, 516, 3354,
  \dodoi{10.1093/mnras/stac2436}

\bibitem[{{Stanton} \& {Murillo}(2016)}]{Stanton2016}
{Stanton}, L.~G., \& {Murillo}, M.~S. 2016, \pre, 93, 043203,
  \dodoi{10.1103/PhysRevE.93.043203}

\bibitem[{{Timmes} \& {Swesty}(2000)}]{Timmes2000}
{Timmes}, F.~X., \& {Swesty}, F.~D. 2000, \apjs, 126, 501,
  \dodoi{10.1086/313304}

\bibitem[{{Trierweiler} {et~al.}(2022){Trierweiler}, {Doyle}, {Melis}, {Walsh},
  \& {Young}}]{Trierweiler2022}
{Trierweiler}, I.~L., {Doyle}, A.~E., {Melis}, C., {Walsh}, K.~J., \& {Young},
  E.~D. 2022, \apj, 936, 30, \dodoi{10.3847/1538-4357/ac86d5}

\bibitem[{{Vanderbosch} {et~al.}(2020){Vanderbosch}, {Hermes}, {Dennihy},
  {Dunlap}, {Izquierdo}, {Tremblay}, {Cho}, {G{\"a}nsicke}, {Toloza}, {Bell},
  {Montgomery}, \& {Winget}}]{Vanderbosch2020}
{Vanderbosch}, Z., {Hermes}, J.~J., {Dennihy}, E., {et~al.} 2020, \apj, 897,
  171, \dodoi{10.3847/1538-4357/ab9649}

\bibitem[{{Vanderbosch} {et~al.}(2021){Vanderbosch}, {Rappaport}, {Guidry},
  {Gary}, {Blouin}, {Kaye}, {Weinberger}, {Melis}, {Klein}, {Zuckerman},
  {Vanderburg}, {Hermes}, {Hegedus}, {Burleigh}, {Sefako}, {Worters}, \&
  {Heintz}}]{Vanderbosch2021}
{Vanderbosch}, Z.~P., {Rappaport}, S., {Guidry}, J.~A., {et~al.} 2021, \apj,
  917, 41, \dodoi{10.3847/1538-4357/ac0822}

\bibitem[{{Vanderburg} {et~al.}(2015){Vanderburg}, {Johnson}, {Rappaport},
  {Bieryla}, {Irwin}, {Lewis}, {Charbonneau}, {Latham}, {Ciardi}, {Schaefer},
  {Kipping}, {Angus}, {Eastman}, {Wright}, {McCrady}, {Wittenmyer}, \&
  {Dufour}}]{Vanderburg2015}
{Vanderburg}, A., {Johnson}, J.~A., {Rappaport}, S., {et~al.} 2015, in
  AAS/Division for Extreme Solar Systems Abstracts, Vol.~47, AAS/Division for
  Extreme Solar Systems Abstracts, 502.02

\bibitem[{{Vauclair} {et~al.}(2015){Vauclair}, {Vauclair}, {Deal}, \&
  {Wachlin}}]{Vauclair2015}
{Vauclair}, S., {Vauclair}, G., {Deal}, M., \& {Wachlin}, F.~C. 2015, in
  Astronomical Society of the Pacific Conference Series, Vol. 493, 19th
  European Workshop on White Dwarfs, ed. P.~{Dufour}, P.~{Bergeron}, \&
  G.~{Fontaine}, 101

\bibitem[{{Veras}(2016)}]{Veras2016}
{Veras}, D. 2016, Royal Society Open Science, 3, 150571,
  \dodoi{10.1098/rsos.150571}

\bibitem[{{Wachlin} {et~al.}(2017){Wachlin}, {Vauclair}, {Vauclair}, \&
  {Althaus}}]{Wachlin2017}
{Wachlin}, F.~C., {Vauclair}, G., {Vauclair}, S., \& {Althaus}, L.~G. 2017,
  \aap, 601, A13, \dodoi{10.1051/0004-6361/201630094}

\bibitem[{{Wachlin} {et~al.}(2022){Wachlin}, {Vauclair}, {Vauclair}, \&
  {Althaus}}]{Wachlin2022}
---. 2022, \aap, 660, A30, \dodoi{10.1051/0004-6361/202142289}

\bibitem[{{Zuckerman} {et~al.}(2011){Zuckerman}, {Koester}, {Dufour}, {Melis},
  {Klein}, \& {Jura}}]{Zuckerman2011}
{Zuckerman}, B., {Koester}, D., {Dufour}, P., {et~al.} 2011, \apj, 739, 101,
  \dodoi{10.1088/0004-637X/739/2/101}

\end{thebibliography}

\end{document}